\documentclass[twocolumn,showpacs,preprintnumbers,amsmath,amssymb]{revtex4}
\usepackage{graphicx}
\usepackage{amsmath}

\begin{document}

\title{ \bf Entangling two superconducting $LC$ coherent modes via a superconducting flux qubit}

\author{\bf Mei-Yu Chen, Matisse W.~Y. Tu and Wei-Min Zhang}
\email{wzhang@mail.ncku.edu.tw} \affiliation{Department of Physics
and Center for Quantum Information Science, National Cheng Kung
University, Tainan 70101, Taiwan } \affiliation{National Center
for Theoretical Science, Tainan 70101, Taiwan}

\begin{abstract}
Based on a pure solid-state device consisting of two superconducting
$LC$ circuits coupled to a superconducting flux qubit, we propose in
this paper that the maximally entangled coherent states of the two
$LC$ modes can be generated for arbitrary coherent states through
flux qubit controls.
\end{abstract}



\maketitle

\section{Introduction}

Quantum entanglement is not only of interests in the fundamentals of
quantum mechanics concerning the EPR paradox \cite{epr}, but also
serves as an indispensable resource for quantum information
processing \cite{NCbook}.  Many discrete entangled states in terms
of polarized photons, atoms, trapped-ions and electrons in
nanostructures have been experimentally demonstrated. However their
practical applications suffer from single-particle decoherence
severely.  Therefore, increasing attention has been paid to
generating macroscopic entangled states \cite{ess,kimbel,maes,duan}
due to their robustness against single-particle decoherence.  The
entangled coherent states is one of the most important ingredients
of quantum information processing using coherent states
\cite{wzhang90}. Creating entangled coherent states, initially
proposed by Sanders in quantum optics \cite{sanders}, have been
extensively explored in many other systems, such as trapped ions
\cite{Munro}, microwave cavity QED \cite{agar}, BEC system
\cite{pan}, as well as the nano-mechanical systems \cite{bose}, but
not yet realized experimentally.

Motivated by the recent experiments on strong coupling between
superconducting $LC$ resonators and superconducting flux qubits
\cite{QLC,b3a}, we propose in this paper a pure electronic
(solid-state) device for generating entangled coherent states of two
superconducting $LC$ modes through flux qubit controls. Using
superconducting qubits coupled with a $LC$ resonator (as a quantum
bus) to generate superconducting qubit couplings, to build two qubit
entanglement, and to implement two-qubit logic gates have been
extensively studied for quantum information processing in the past
years \cite{Yu99,You02,Zhou04,Mig05,Liu06,Zag06,Spi06,Bl07,
circuitQED1,circuitQED2,hime06}. Here, we shall design an
alternative superconducting circuits that using the measurement of
superconducting flux qubit states to generate the maximum
entanglement states of the two $LC$ coherent modes for quantum
communication. The scheme of generating entanglement states of
distant systems through measurement was indeed proposed a decade ago
\cite{PRA59-1025}. However, $LC$ circuits are building blocks of all
the electronic information and communication devices used today,
this entangled $LC$ coherent mode generator could be very promising
for practical realization of quantum communication and quantum
information processing.

\section{System setup}

The device we design here consists of two superconducting $LC$
circuits strongly coupled to a superconducting flux qubit.  Fig.~1A
is a schematic setup of our superconducting circuits. The central
circuit is a superconducting flux qubit which is coupled to two
superconducting $LC$ circuits through mutual inductance. The qubit
is enclosed by a superconducting quantum interference device (SQUID)
as a qubit measurement device. Coherent control of the qubit is
achieved via two microwave control lines ($I_1, I_2$).  Symmetric
circuits are designed to suppress excitation of the SQUID and to
protect the two $LC$ oscillators from the unwanted influence of the
qubit controlling pulses.

\begin{figure}[h]
\centering
\includegraphics[width=8cm,height=7.5cm]{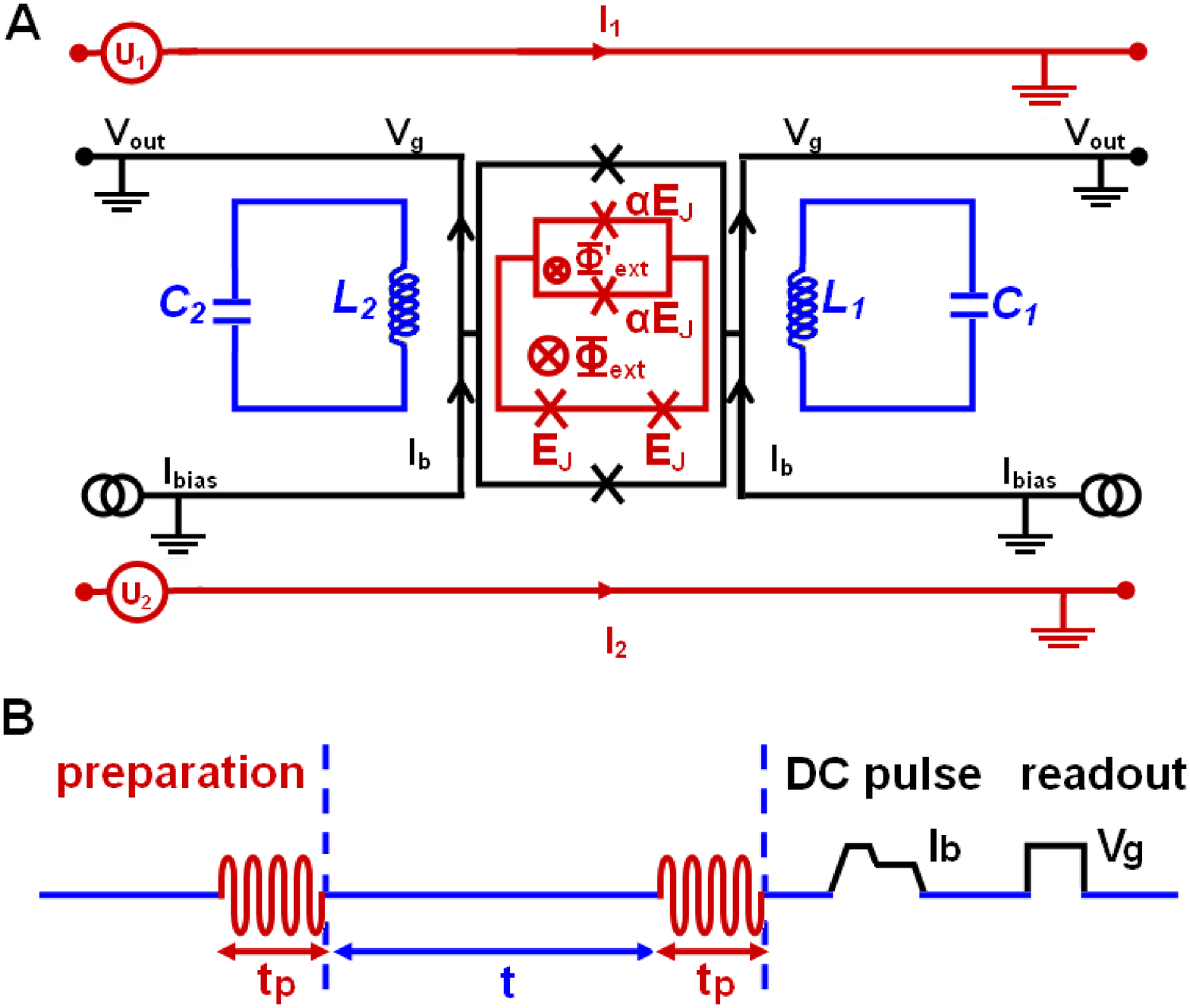}
 \caption{ (A). A schematic diagram of the pure electronic device
for entangling two $LC$ modes through a flux qubit. The four
junctions flux qubit is in the inner loop, and is enclosed by a
dc-SQUID detector (with two Josephson junctions). The two microwave
lines modulate the flux in the qubit loop, and control the
parameters $\Delta$ and $\varepsilon$.  The qubit state are read out
by applying a current pulse $I_b$ and then recording the voltage
state of the SQUID. (B). Signals involved in quantum state
manipulation and measurement. First, microwave pulses are applied to
the qubit for state preparation. After the last microwave pulse, a
readout current pulse $I_b$ is injecting to the dc-SQUID. The height
and the length of the pulse are adjusted to give the best
discrimination between the ground and the excited state. Finally,
measuring the voltage state of the dc-SQUID in which the voltage
state of the dc-SQUID depends on the switching probability of the
energy eigenstates. }
\end{figure}

Both the superconducting LC circuits and the flux qubit can be
fabricated on a chip down to the micrometer scale.  The
superconducting $LC$ circuit is an ideal harmonic oscillator
verified experimentally \cite{QLC}, and the two levels of the
superconducting flux qubit comprise of the clockwise  and
counterclockwise persistent-current states $|0\rangle$ and
$|1\rangle$ \cite{285,299}.  The latter is made of a superconducting
loop interrupted by three Josephson junctions \cite{285} in which
two junctions have the same Josephson coupling energy $E_J$, and the
third junction (placed by a SQUID in Fig.~1A) has the coupling
energy smaller than that of the other two junctions by a factor
$\alpha$ with $0.5<\alpha<1$. The interaction of the flux qubit and
two $LC$ circuits can be controlled by the external microwave
control lines. The geometrical structure of the $LC$ circuit is
adjustable so that the strong coupling can be achieved \cite{QLC}.
The flux qubit is also tunable and has the advantage of
long-decoherence time. These advantages decrease the difficulty of
the experiment and increase the feasibility.

Preparing the flux qubit in a superposition of the states
$|0\rangle$ and $|1\rangle$  initially, we are able to drive the
qubit and the two $LC$ modes into a tripartite entanglement [see
Fig.~1B and Eq.~(\ref{tri})]. Measuring the qubit state with an
enclosed dc-SQUID, which is inductively coupled to the qubit
\cite{299,296} as shown in Fig.~1A, will generate the entangled $LC$
coherent modes.  This is the procedure of entangling two
superconducting $LC$ coherent modes through flux qubit controls. As
schematically depicted in Fig.~1A, the qubit detector consists of a
ring interrupted by two Josephson junctions. This SQUID is connected
in such a way that the current can be injected through the parallel
junctions. The switching current of the detector is sensitive to the
flux produced by the current of the flux qubit. The readout of the
qubit state is performed by applying a pulse sequence to the SQUID,
as shown in Fig.~1B, and recording whether the SQUID had been
switched to a finite voltage ($V_g$) or remained in the zero
voltage.

\section{Entangling two $LC$ coherent modes}

Explicitly, the Hamiltonian of the total system can be described by
\cite{QLC}
\begin{align}
H=\sum_{i=1}^2\hbar\omega_ia^\dag_i a_i
-\hbar\Big({\varepsilon\over2}\sigma_z+{\Delta\over2}\sigma_x\Big)
+\sum_{i=1}^2 \hbar\lambda_i(a^\dag_i+a_i)\sigma_z, \label{mh}
\end{align}
where $a^\dag_i (a_i), i=1,2$ is the plasmon creation (annihilation)
operator of the two $LC$ oscillators, the corresponding resonance
frequency $\omega_i$ is determined by the respective capacitance
$C_i$ and the inductance $L_i$: $\omega_i={1\over\sqrt{L_iC_i}}$
which is of the order of tens GHz for a micrometer scale $LC$
circuit. The operators $\sigma_z$, $\sigma_x$ are the usual Pauli
matrices describing the superconducting flux qubit. The energy
splitting of the qubit is given by
$\hbar\varepsilon=2I_p(\Phi_{ext}-{\Phi_0\over2})$ in which $I_p$ (
0.3$\sim$0.5 $\mu$A) is the persistent current in the qubit,
$\Phi_{ext}$ is the external magnetic flux applied in the
superconducting loop and $\Phi_0={h\over2e}$ is the flux quantum.
$\Delta$ is an effective tunneling amplitude describing qubit state
flip, which depends on $E_J$ \cite{431}. The Josephson energy $E_J$,
in turn, can be controlled when the third junction is replaced by a
SQUID, as shown in Fig.~1A, introducing the flux $\Phi'_{ext}$ as
another control parameter \cite{285}. These two external magnetic
flux $\Phi_{ext}$ and $\Phi'_{ext}$ can be suddenly switched by two
resonant microwave lines $I_1$ and $I_2$ for a finite time ($\sim$
tens of ps) to manipulate the two parameters, $\varepsilon$ and
$\Delta$, respectively \cite{87}. The $LC$ circuits couple to the
flux qubit via the mutual inductance with the coupling constant
$\lambda_i=M_iI_p\sqrt{{\omega_i\over2\hbar L_i}}$, where $M_i$
($\sim$pF) is the mutual inductance between the $LC$ circuits and
the flux qubit \cite{93A,QLC}. As we can see, the qubit energy
splitting $\varepsilon$ and the $LC$-qubit coupling $\lambda_i$ are
related through the persistent current $I_p$ while the qubit flip
amplitude $\Delta$ can be almost independently controlled through
the additional external flux $\Phi'_{ext}$.

The manipulating and measuring signal sequences on the flux qubit
are shown in Fig.~1B. First let the $LC$ circuits be prepared in
their ground states and the flux qubit in the state $|0\rangle$, the
state of the total system at $t=0$ can then be written as
$|\Psi(0)\rangle=|0\rangle|0_10_2\rangle$ where the subscripts $1$
and $2$ denote the two $LC$ circuits. The qubit localized in
$|0\rangle$ at $t=0$ also implies that the qubit flip amplitude
$\Delta$ is initially adjusted to be much smaller ($\approx 0$) in
comparing with the values of $\varepsilon$ and $\lambda_i$, i.e.
$\Delta_0 \ll \varepsilon, \lambda_i $. Then applying a pulse
non-adiabatically [denoted by $P(t_p)$] to modulate the two control
lines $I_1$ and $I_2$ such that $\Phi_{ext}$ is kept almost no
change but $\Phi'_{ext}$ is changed dramatically. Since the $LC$
resonators couple to the flux qubit through the $\sigma_z$ component
in our device \cite{QLC,new08,legt}, the $LC$-qubit coupling
$\lambda_i$ is only sensitive to the change of the energy splitting
$\varepsilon$ of the qubit [also see the explicit expressions given
after Eq.(\ref{mh})]. This allows us to keep the parameters
$\varepsilon, \lambda_i$ almost as a constant but adjust the qubit
flip amplitude $\Delta$ quickly to a large value to reach a
condition $\Delta \gg \varepsilon, \lambda$ through the
non-adiabatical change of $\Phi'_{ext}$. As a result, this pulse
drives the flux qubit into the degeneracy point within a duration
$t_p=\frac{\pi}{2\Delta}$ without disturbing the $LC$ resonator
states too much. Accordingly, the state $|\Psi(0)\rangle$ evolves to
\begin{align}
|\Psi(t_p)\rangle=
{1\over\sqrt{2}}(|0\rangle+i|1\rangle)|0_10_2\rangle .
\end{align}
In fact, it has been shown recently that for a similar system
\cite{b3a,Koch,new08}, the qubit flip amplitude $\Delta$ can be
rapidly increased while the qubit energy splitting $\varepsilon$ and
the qubit-$LC$ coupling $\lambda$ vanishes rather abruptly through
the non-adiabatical control of the flux (see explicitly Fig.~3 in
\cite{new08}).

After the first pulse, the parameters return to the initial values
$\Delta \rightarrow \Delta_0 \approx 0$, namely the $\sigma_x$ term
in the Hamiltonian now contributes little effect on the subsequent
evolution of the qubit. Then let the system evolve lasting a period
of time $t$, the resulted state is given by
\begin{align}
|\Psi(t_p+t)\rangle={1\over\sqrt{2}}\big[&e^{-{i\epsilon
t\over2}}|0\rangle|\kappa_1(t)\kappa_2(t) \rangle \nonumber
\\ & +ie^{{i\epsilon t\over 2}}
|1\rangle|-\kappa_1(t)-\kappa_2(t)\rangle\big] \label{E5}
\end{align}
where $|\kappa_i(t)\rangle\equiv
e^{\kappa_{i}(t)a_{i}^{\dag}-\kappa_{i}^{*}(t)a_{i}}|0_i\rangle$ is
a coherent state characterized by the complex variable
$\kappa_i(t)={\lambda_i\over\omega_i}(1-e^{-i\omega_it})$. Equation
(\ref{E5}) is a tripartite entangled state of one qubit with two
coherent $LC$ modes. We can apply the same pulse $P(t_p)$ to the
flux qubit again (see Fig.~1B), the state of Eq.~(\ref{E5}) is
driven to
\begin{align}
|\Psi&(t_p+t+t_p)\rangle = \nonumber \\
&{1\over2}e^{-{i\varepsilon t\over 2}}
\Big[|0\rangle\big(|\kappa'_1(t)\kappa'_2(t)\rangle-e^{i\varepsilon
t}|-\kappa'_1(t)-\kappa'_2(t)\rangle\big)\nonumber\\
&~~~~~+i|1\rangle\big(|\kappa'_1(t)\kappa'_2(t)\rangle+e^{i\varepsilon
t}|-\kappa'_1(t)-\kappa'_2(t)\rangle\big)\Big] \label{tri},
\end{align}
where $\kappa'_i(t)= \kappa_i(t) e^{-i\omega_i t_p}$. We now measure
the flux qubit in the $\sigma_z$ basis, i.e. the natural
computational basis \{$|0\rangle$, $|1\rangle$\} which is indeed the
energy eigenstate basis in the present case since $\Delta
\rightarrow \Delta_0 \ll \varepsilon $ after the second pulse. As a
result, the two $LC$ modes collapse into the state:
\begin{align}
|\psi_+\rangle_{12}={1\over\sqrt{2}}\big[|\kappa'_1(t)
\kappa'_2(t)\rangle +e^{i\varepsilon t}
|-\kappa'_1(t)-\kappa'_2(t)\rangle\big] \label{ecs1}
\end{align}
if the qubit is measured with the result $1$, or
\begin{align}
|\psi_-\rangle_{12}={1\over\sqrt{2}}(\big[|\kappa'_1(t)
\kappa'_2(t)\rangle-e^{i\varepsilon t}
|-\kappa'_1(t)-\kappa'_2(t)\rangle\big] \label{ecs2}
\end{align}\\
if the measured result is $0$. Each outcome has a probability of
$50\%$ to occur. Eqs.~(\ref{ecs1}-\ref{ecs2}) are two entangled
coherent states of the two superconducting $LC$ circuits  we propose
to generate.

In practice, we are more interested in the case of the two
superconducting $LC$ circuits being symmetric in geometry for
protecting the two $LC$ oscillators from the unwanted influence of
the qubit controlling pulses. Thus we have $\omega_1=\omega_2 \equiv
\omega$ and $\lambda_1=\lambda_2 \equiv \lambda$. Let the system
evolve for a period of time $t= \frac{\pi}{\omega}$ between the two
pulses, we obtain the following standard form of the entangled two
$LC$ coherent states
\begin{align}
|\phi_\pm \rangle_{12} = {1\over\sqrt{2}}
\big[|2\kappa_02\kappa_0\rangle \pm
e^{i\varphi}|-2\kappa_0-2\kappa_0\rangle \big]  \label{ecs0}
\end{align}
with $\kappa_0={\lambda\over\omega}e^{-i\omega t_p}$ and
$\varphi=\pi\frac{\varepsilon}{\omega}$. Using the concept of
concurrence for bipartite entangled non-orthogonal states
\cite{Munro,Wang02}, it is easy to show that the concurrence for
$|\phi_\pm \rangle_{12}$ is given by
\begin{align}
  C_{\phi_\pm}=\frac{1-e^{-16|\kappa_0|^2}}{1 \pm
  e^{-16|\kappa_0|^2}\cos\varphi}
\end{align}
If the rate $|\kappa_0|=\frac{\lambda}{\omega} \ge 0.5$, the
exponential factor $e^{-16\kappa_0^2}\ll 1$. Then we have
$C_{\phi_\pm}\simeq 1$, namely, $|\phi_\pm \rangle_{12}$ are nearly
maximally entangled even though the average boson number
($=4|\kappa_0|^2$) in the coherent state $|2\kappa_0\rangle$ is a
small number. By well-designed circuits, one can let the ratio of
coupling constant to the resonance frequency near to one, i.e.
$\kappa_0\simeq 1$, then $|\langle-2\kappa_0|2\kappa_0\rangle|^2
=e^{-16|\kappa_0|^2}\simeq10^{-7}\simeq0$, namely, the two coherent
states $|-2\kappa_0\rangle$ and $|2\kappa_0\rangle$ in the entangled
state (\ref{ecs0}) can be nearly orthogonal.

However, if the average boson number is too small, the coherent
states $|\kappa_i(t)\rangle$ are not truly macroscopic states such
that the robustness against decoherence for the corresponding
entanglement states could be faded. This weakness can be overcome by
preparing the two $LC$ circuits initially in two coherent states
$|\alpha_1\rangle$ and $|\alpha_2\rangle$, while the flux qubit is
still in the ground state $|0\rangle$. The initial state of the
total system becomes $|\Psi'(0)\rangle=|0\rangle|\alpha_1 \alpha_2
\rangle$. Similarly using the pulse $P(t_p)$ to rotate the qubit
state: $|\Psi'(t_p)\rangle={1\over\sqrt{2}} (|0\rangle+i|1\rangle)
|\alpha'_1 \alpha'_2\rangle$ where
$\alpha'_i=\alpha_ie^{-i\omega_it_p}$. Then let the system evolve
for a period of time $t$, the resulting state of the total system
is:
\begin{align}
|\Psi'(t_p+t)\rangle={1\over\sqrt{2}}&
\big[e^{-i\theta}|0\rangle|\beta_{1+}(t)\beta_{2+}(t) \rangle\nonumber\\
& + i e^{i\theta}|1\rangle|\beta_{1-}(t)\beta_{2-}(t)\rangle\big].
\end{align}
Here we have defined $|\beta_{i\pm}(t)\rangle \equiv |\alpha'_i
e^{-i\omega_i t}\pm \kappa_i(t)\rangle$ and $\theta = {\varepsilon
\over 2}t+i(\delta_1+\delta_2)$ with $\delta_i \equiv
\frac{\lambda_i}{2\omega_i}[(e^{i\omega_it}-1){\alpha'}^*_i
+(1-e^{-i\omega_it})\alpha'_i]$. Again we can measure the flux qubit
in the $\sigma_z$ basis after reapplying the pulse $P(t_p)$ to the
qubit, which results in the following entangled coherent states,
\begin{align}
|\psi'_\pm
\rangle_{12}=\frac{1}{\sqrt{2}}\big[|\beta'_{1+}(t)\beta'_{2+}(t)\rangle
\pm e^{i2\theta}| \beta'_{1-}(t)\beta'_{2-}(t)\rangle\big],
\end{align}
where $\beta'_{i\pm}=\beta_{i\pm}e^{-i\omega_it_p}$. Now, the
coherent states $|\beta'_{i\pm}(t)\rangle$ can be a very macroscopic
state, depending on the initial voltages applied to the two $LC$
circuits for generating the initial two coherent states
$|\alpha_i\rangle$. While the entanglement measures of $|\psi'_\pm
\rangle_{12}$ are almost the same as that of $|\psi_\pm
\rangle_{12}$. To be specific, we consider the symmetric $LC$
circuits again with $\alpha_1=\alpha_2\equiv \alpha$ and take
$t=\frac{\pi}{\omega}$, the concurrence for $|\psi'_\pm
\rangle_{12}$ is given by
\begin{align}
  C_{\psi'_\pm}=\frac{1-e^{-16|\kappa_0|^2}}{1 \pm
  e^{-16|\kappa_0|^2}\cos(\varphi-16|\kappa_0|{\rm Im}
  \alpha' )}.
\end{align}
It shows that for a given $|\kappa_0|=\frac{\lambda}{\omega}$ such
that $e^{-16|\kappa_0|^2}\ll 1$, we can always have
$C_{\psi'_\pm}\simeq 1$ regardless of the value of $\alpha'=\alpha
e^{i\omega t_p}$. In other words, the device we proposed here can
generate maximally entangled states for arbitrary coherent states
with arbitrary large oscillating amplitudes.

\section{decoherence analysis and conclusion}

We have shown how to entangle two $LC$ coherent modes through a
superconducting flux qubit. To make the device feasible, we should
also analyze various possible decoherence effect to the system. In
solid-state systems decoherence comes from many redundant degrees of
freedom that interact with the device. The noise may due to the
emission from the superconducting $LC$ circuits and the flux qubit,
and from the control and detect of the qubit state. (i) In fact, the
decoherence of the entangled coherent state due to the photo loss
has been analyzed in detail recently by one of us in
\cite{wzhang07}. (ii) Recent experiments demonstrated that the
relaxation and dephasing times of the flux qubit are greater than
0.1$\mu$s \cite{b3a,hime06,b3b,b3c}, longer enough for qubit
operations which is of the order of tens of picoseconds, estimated
from $t_p =\frac{\pi}{2\Delta}\sim 40$ ps for $\Delta \simeq 40 $
GHz \cite{431}. (iii) The SQUID may be inductively coupled to the
two $LC$ oscillators. But from the estimation of the Johnson-Nyquist
noise in the bias circuit, it has been shown that this contribution
is several orders of magnitude weaker \cite{JNnoise,QLC}. (iv) The
symmetric design of the $LC$ as well as the dc-SQUID circuits has
effectively suppressed the noise induced by qubit operations
\cite{QLC,hime06,299,431}. Put all these decoherence effects
together, the estimated decoherence times from the different source
are much longer than the typical time scale (the pulse time $t_p
\sim 40 $ ps and evolving time $t=\frac{\pi}{\omega} \sim 0.1$ ns
for $\omega \simeq 40$ GHz \cite{QLC}) of the system for producing
entangled coherent states, which makes the system more practical.

In conclusion, we proposed a pure electronic (solid state) device
consisting of two superconducting $LC$ modes coupled with a
superconducting flux qubit. We showed that entangled coherent states
of the two $LC$ modes can be generated through the flux qubit
controls. With the well-designed superconducting circuits one can
achieve a strong coupling between the flux qubit and the $LC$
circuits \cite{QLC,b3a}, and the adjustable physical parameters
gives extra degrees of freedom to generate the maximally entangled
states for arbitrary coherent states. Beside being of the
fundamental interest, the robust, macroscopic entanglement of two
$LC$ coherent modes described here is expected to be useful and
powerful in quantum information processing. Such an entangled
coherent state generating device is promising in practical
applications since $LC$ circuits are the building blocks of the
information technology. Once the entanglement coherent states of the
two $LC$ modes can be experimentally realized, it is easy to create
quantum channels by emitting one of the entangled $LC$ modes to a
receiver at a long distance (see a schematic plot in Fig.~2). These
features make this entangled coherent state generator unique in the
further development of quantum information processing. Finally, the
$LC$ circuits coupling to a flux qubit on a chip and the operations
and detection of the qubit states require no new technology as far
as we can see, as all the essential techniques have already been
developed in various experiments. These advantages increase the
feasibility of this entanglement coherent state generator in
practice.

We would like to thank Dr. Mariantoni for bring our attention to
their recent work on a similar system with different motivation
\cite{Mar08} after we completed this paper. This work is supported
by the National Science Council of ROC under Contract
No.~NSC-96-2112-M-006-011-MY3.

\begin{figure}[h]
\includegraphics[width=9cm,height=5cm]{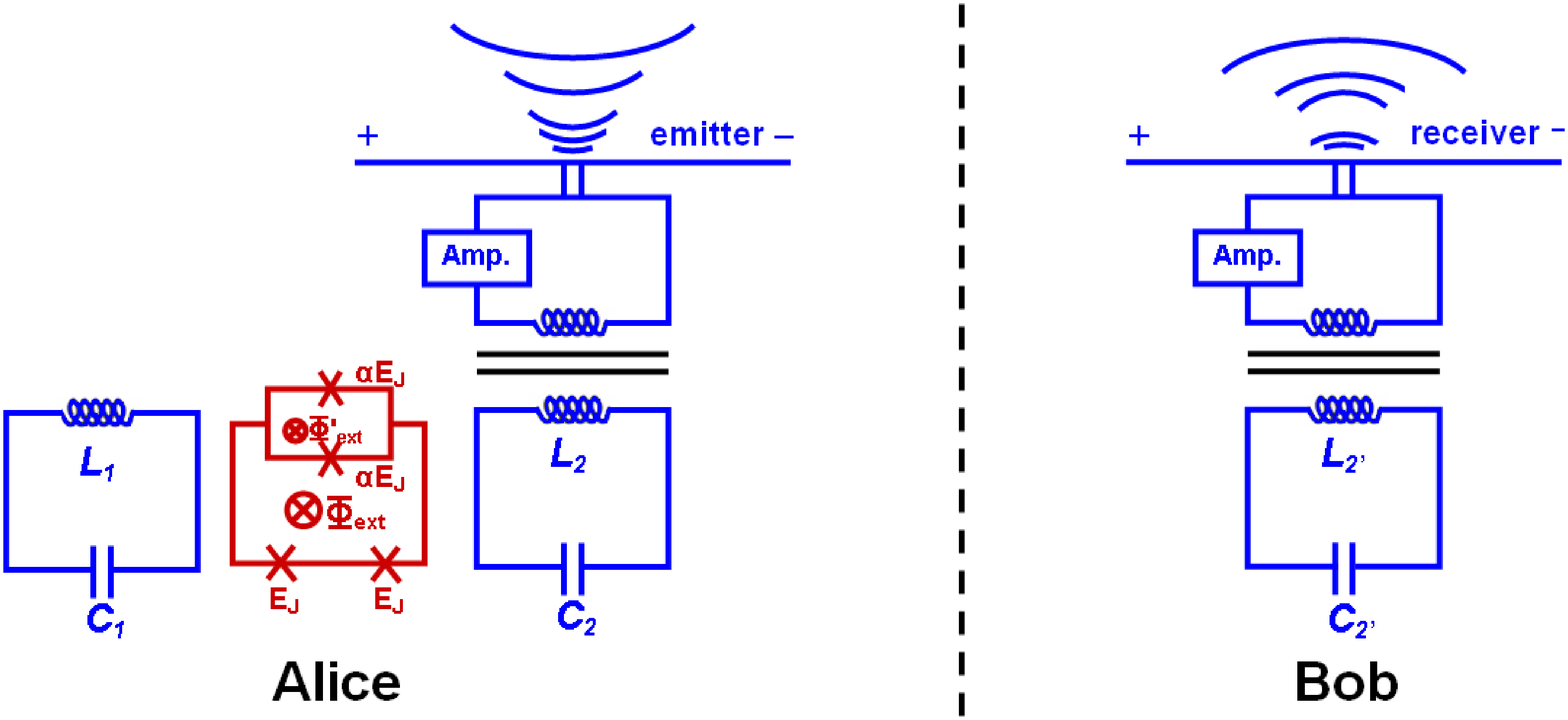}
 \caption{ After generating the entangled coherent state of the two $LC$ modes,
  we can use the antennas to emit one of the two entangled $LC$ modes
  to a long distance receiver without using waveguides or fibers.}
\end{figure}

\end{document}